\documentclass[twocolumn,showpacs,amsmath,amssymb, superscriptaddress]{revtex4}

\usepackage{amsmath}
\usepackage{amssymb}
\usepackage{amsfonts}

\usepackage{graphicx}
\usepackage{float}
\usepackage{color}

\usepackage{amsthm}

\DeclareMathOperator{\Tr}{Tr}

\newcommand{\angl}[1]{{\left\langle #1 \right\rangle}}

\makeatletter
\def\loweq@align#1#2{\lower.6ex\vbox{\baselineskip\z@skip\lineskip\z@
    \ialign{$\m@th#1\hfil##\hfil$\crcr#2\crcr=\crcr}}}
\def\lowsim@align#1#2{\lower.6ex\vbox{\baselineskip\z@skip\lineskip\z@
    \ialign{$\m@th#1\hfil##\hfil$\crcr#2\crcr\sim\crcr}}}
\def\geqq{\mathrel{\mathpalette\loweq@align >}}
\def\leqq{\mathrel{\mathpalette\loweq@align <}}
\def\grsim{\mathrel{\mathpalette\lowsim@align >}}
\def\lesssim{\mathrel{\mathpalette\lowsim@align <}}
\def\gsim{\mathrel{\mathpalette\lowsim@align >}}
\def\lsim{\mathrel{\mathpalette\lowsim@align <}}
\makeatother
\newcommand{\grless} 
{ {\, \raise-.24em\hbox{$<$} \hspace{-0.8em} \raise.31em\hbox{$>$}\, } }
\newcommand{\lessgr} 
{ {\, \raise-.24em\hbox{$>$} \hspace{-0.8em} \raise.31em\hbox{$<$}\, } }
\newfont{\bg}{cmr10 scaled\magstep4}                    
\newcommand{\bigzerou}{\smash{\lower1.7ex\hbox{\bg 0}}}

\newcommand{\crl}[1]{[-\infty,\infty]}

\newcommand{\Ref}[1]{(\ref{#1})}

\newcommand{\da}[1]{#1^\dag}

\newcommand{\av}[1]{\langle#1\rangle}

\begin{document}

\title{Time-Optimal Transfer of Coherence}
  \author{Alberto Carlini}
 \email{acarlini@mfn.unipmn.it}
 \affiliation{Dipartimento di Scienze ed Innovazione Tecnologica, Universita' del Piemonte Orientale, Alessandria, Italy}
 \affiliation{Istituto Nazionale di Fisica Nucleare, Sezione di Torino, Gruppo Collegato di Alessandria, Italy}

\author{Tatsuhiko Koike}
 \email{koike@phys.keio.ac.jp}
 \affiliation{Department of Physics, Keio University, Yokohama, Japan}

\date{November 10, 2012}
\begin{abstract}
We provide exact analytical solutions for the problem of time-optimal transfer of coherence from one spin polarization to a three-fold coherence in a trilinear Ising chain with a fixed energy available and subject to local controls with a non negligible time cost.
The time of transfer is optimal and consistent with a previous numerical result obtained assuming instantaneous local controls.

\end{abstract}  


\maketitle

\section{Introduction}

The techniques of quantum optimal control are ubiquitous and have multiple applications in, e.g., molecular processes \cite{shapiro}, nuclear magnetic resonance (NMR) spectroscopy \cite{cavanagh}, quantum information theory 
\cite{chuangnielsen}-\cite{brif}.
In particular, time-optimal control, where the goal is to minimize the physical time to reach a quantum target 
(i.e. a quantum state or a unitary operation) is important for the construction of efficient gates in quantum computing architectures and it offers a more physical framework in defining the complexity of quantum algorithms \cite{schulte}.
On the other hand, controlling the spin dynamics in quantum chains is of relevance in the efficient exchange of quantum information \cite{murphy}, and, e.g., in NMR multidimensional spectroscopy experiments.
The problem of the efficient transfer of coherence in a trilinear spin chain where neighbor spins are subject to equal 
Ising couplings and where the single qubits are separately addressed via instantaneous, local controls was 
discussed within a geometrical quantum control ansatz in \cite{khaneja2}.
The case of unequal Ising couplings and to the case of longer chains was treated in 
\cite{yuan}-\cite{nimbalkar}.
The time duration of the transfer given numerically in \cite{khaneja2}, though shorter than the one obtained with conventional approaches, was not guaranteed to be a (global) time-optimal control. 
One of the postulates in the above works is that the time cost of one-qubit operations is zero,  
which is reasonable (e.g., in heteronuclear NMR) when the other time-scales (e.g., the inverse of the maximal Ising couplings) involved in the quantum control process are much longer.
In a series of papers \cite{pure}-\cite{unitary}, we discussed a theoretical scheme for time-optimal quantum control,
which we called the quantum brachistochrone (QB \cite{greek}), 
where the postulate of zero-time local operations is not required.
The scheme was 
The QB is derived from an action principle which enforces the dynamical laws of quantum evolution (i.e., the
Schr\"odinger equation or a master equation) and the constraints which the Hamiltonian of the physical system 
has to satisfy (e.g., a fixed total energy, the absence of certain qubit interactions etc..).
The framework was designed for the time-optimal evolution of quantum states between fixed initial and final states
\cite{pure}-\cite{mixed}, for the time-optimal generation of a certain unitary quantum gate \cite{unitary}, and for the
situation (typical in experiments) where the target is reachable only in an approximate way (i.e., with a fidelity 
smaller than one \cite{fidelity}).

In this brief note we reconsider the problem of the time-optimal transfer of one qubit polarization to three 
qubit quantum coherence studied in \cite{khaneja2}.
We use a trilinear qubit system with the same Ising interaction Hamiltonian as in \cite{khaneja2}, but
we consider a generic and time consuming local control on the intermediate qubit of the chain and 
we assume that a finite energy is available.
By using the QB methods, we are able to analytically quantify the shape of the local control and the duration of the optimal transfer, which essentially coincides with the numerical value found in \cite{khaneja2}.


\section{Quantum Brachistochrone}

Let us first review the main formalism of the QB as discussed, e.g., in \cite{unitary}.
The problem of the QB for unitary operations is to find the time-optimal way to generate a target quantum gate $U_f$
via the control of a Hamiltonian $H(t)$.
We impose that the quantum evolution is driven by the Schr\"odinger equation, and that the Hamiltonian has to
satisfy certain constraints.
For example, the energy available in the experiments may be limited, or some interactions between the qubits may be forbidden.
The QB problem can be concisely formulated in terms of the action \cite{mixed,unitary}:
\begin{align}
  \label{eq-action}
  S(U,H; \alpha, \Lambda,\lambda_j) &:=\int_0^1 ds \left[ N\alpha +L_S+L_C\right],
  \\
  \label{eq-LS}
  L_S &:=\av{\Lambda,i \tfrac{dU}{ds}\da U- \alpha H},
  \\
  \label{eq-LC}
  L_C &:= \alpha   \sum_j{\lambda_j}f^j(H), 
\end{align}
where $\angl{A,B}:=\Tr (\da AB)$ and $\log N$ is the number of qubits.
The Hermitian operator $\Lambda(s)$ and the real functions $\lambda_j(s)$ are Lagrange multipliers. 
The function $\alpha (s)$ gives the time cost and it relates the physical $t$ and the parameter $s$ times
via $t:=\int \alpha(s) ds$ \cite{mixed}.

The multiplier $\Lambda$ enforces the Schr\"odinger equation:
\begin{align}
  \label{eq-Sch}
  i\frac{dU}{dt}=HU.
\end{align}

The constraints on the Hamiltonian are obtained from the variation of the action \Ref{eq-action}
with respect to $\lambda_j$:
\begin{align}
  f_j(H)=0. 
\label{constraints}  
\end{align}
For instance, the finite energy condition reads: 
\begin{align}
f_0(H) :=\tfrac{1}{2}[\Tr(H^2)- N\omega^2]=0,
\label{eq-normH}
\end{align} 
with $\omega := \mathrm{const}$.

The variation of $S$ with respect to $H$ gives:
\begin{align}
\Lambda =  \lambda_0 H + \sum_{j\not = 0} \lambda_j \frac{\partial f_j(H)}{\partial H},
\label{eq-H}
\end{align} 
while the variation with respect to $\alpha$ gives a normalization condition for $\Lambda$:
\begin{align}
\Tr (H\Lambda)=N.
\label{eq-alpha}
\end{align}

Finally, variation of the action with respect to $U$ gives, after some elementary steps:
\begin{align}
  \label{eq-fund}
  i\frac{d\Lambda}{dt}= [H, \Lambda].
\end{align}
We name eq. \Ref{eq-fund} the {\it quantum brachistochrone} equation.
Together with the constraints \Ref{constraints} it defines a boundary-value problem for the evolution operator
$U(t)$ with fixed initial ($U(t=0)=1$) and final conditions ($U(t=T)=U_f$), where $T$ is the optimal time duration 
of the quantum evolution.
Once the target quantum gate $U_f$ is given, one can solve the quantum brachistochrone \Ref{eq-fund} 
together with the constraints \Ref{constraints} and find the time-optimal Hamiltonian $H_{\mathrm{opt}}(t)$, which
will depend upon a set of integration constants.
Then, integration of the Schr\"odinger equation \Ref{eq-Sch} from $U(0)=1$ produces the time-optimal evolution
$U_{\mathrm{opt}}(t)$.
Finally, the integration constants (including the time duration $T$) in $H_{\mathrm{opt}}(t)$ are determined
imposing that $U_{\mathrm{opt}}(T)=U_f$ \cite{unitary}.

\section{Ising Hamiltonian}

Let us now explicitly use the QB formalism for the model of a linear Ising chain of three qubits (identified by
indices $a\in \{1, 2,  3\}$) with Ising couplings $J_{12}$ and $J_{23}$.
We further assume that the intermediate qubit of the chain is controlled via a local magnetic field 
$B_i(t)$ ($i=x, y, z$), i.e., we consider the Hamiltonian:
\begin{eqnarray}
 H(t):=J_{12}\sigma_z^{1}\sigma_z^{2} +J_{23}\sigma_z^{2}\sigma_z^{3} +\vec{B}(t)\cdot \vec{\sigma}^{2},
\label{ising}
\end{eqnarray}
where, e.g., $\sigma_i^1\sigma_j^2:=\sigma_i\otimes \sigma_j\otimes 1$,
and $\sigma_i$ are the Pauli
operators \cite{chuangnielsen}.
The interaction part in eq. \Ref{ising} is the same as $H_c$ in eq. (2) of \cite{khaneja2}.
However, the local and time consuming term $\vec{B}(t)\cdot \vec{\sigma}^{2}$ is used instead of the zero cost 
terms $H_A$ and $H_B$ of \cite{khaneja2}.
We also introduce the couplings' ratio $K:=J_{23}/J_{12}$, rescale time as $\tau :=J_{12} t$, the magnetic 
field as $\hat B(\tau) :=B(t)/J_{12}$ and the energy as $\hat\omega := \omega /J_{12}$. Then, e.g., 
the finite energy condition \Ref{eq-normH} explicitly reads:
\begin{equation}
{\vec{\hat B}}^2=\hat \omega^2 -(1+K^2)=\mathrm{const}.
 \label{const-H}
\end{equation}

The analytical solution of the quantum brachistochrone problem 
for the Hamiltonian \Ref{ising} was found in Ref. \cite{fidelity} (for details, se Sections III and VIIIA there).
Here we are interested in one of the main results, i.e. the time-optimal magnetic field which reads:
\begin{align}
  {\vec{\hat B}}_{\mathrm{opt}}(\tau)= \left ( \begin{array}{c}
 {\hat B_0}\cos \theta(\tau)\\
 {\hat B_0}\sin \theta(\tau)\\
   \hat B_z   
  \end{array}\right ),
  \label{bopt}
\end{align}
where $\theta(\tau):=\hat \Omega \tau +\theta_0$ and $\hat \Omega$, $\hat B_z$, $\hat B_0$ and $\theta_0$ are integration constants.

The explicit form of the time-optimal evolution operator $U_{\mathrm{opt}}(t)$, which in principle can be obtained
integrating the Schr\"odinger eq. \Ref{eq-Sch} and using eqs. \Ref{ising} and \Ref{bopt}, is not necessary
for the purposes of this paper (for details see \cite{ising}).

\section{Time-Optimal Transfer of Coherence}

Our goal now is to time-optimally create a 3-qubit quantum coherence starting from an initial polarization on one of the qubits at the ends of the Ising chain.
For instance, we would like to realize the following transformation:
\begin{align}
\sigma^1_x \longrightarrow ~\sigma^1_y \sigma^2_y \sigma^3_z, 
\label{transfer}
\end{align}
and find the time-optimal control of the Hamiltonian \Ref{ising} which performs this transformation.
We call this as the time-optimal coherence transfer (TOCT) problem.

In the previous literature \cite{khaneja2}, the TOCT problem has been addressed in the following way.
First, one introduces expectation values of operators $\langle O\rangle(\tau)  := \Tr [O \rho(\tau)]$, 
where the density operator $\rho(\tau)=U(\tau)\rho(0)\da U(\tau)$ and $U(\tau)$ satisfies the 
Schr\"odinger equation \Ref{eq-Sch} with the Hamiltonian \Ref{ising}.
Then, one notes that the algebra generated by the Hamiltonian \Ref{ising} and the operator $\sigma^1_x$ is just an 8-d  subspace of the full 63-d algebra for the three qubits, and is spanned by
$x_1:=\langle \sigma_x^1\rangle$, $~x_2:=\langle \sigma_y^1\sigma_x^2\rangle$ , $~x_3:=\langle \sigma_y^1\sigma_z^2\rangle$,
$~x_4:=\langle \sigma_y^1\sigma_y^2\rangle$, $~x_5:=\langle \sigma_x^1\sigma_z^3\rangle$, $~x_6:=\langle \sigma_y^1\sigma_x^2\sigma_z^3\rangle$, $~x_7:=\langle \sigma_y^1\sigma_z^2\sigma_z^3\rangle$ and $~x_8:=\langle \sigma_y^1\sigma_y^2\sigma_z^3\rangle$.     
Using $\langle O\rangle^\cdot =\Tr[O\dot \rho]=i\langle [H, O] \rangle$ one 
gets a system of equations for the $x_i$s, which depend upon the unknown control $\vec{\hat B}(\tau)$ in \Ref{ising}. 
Then one should solve such a system and determine the optimal control $\vec{\hat B}_{\mathrm{opt}}(\tau)$
ensuring that the transfer from the initial condition $x_1(0)=1$ to the final condition $x_8(\tau_\ast)=1$ happens in 
the minimal time $\tau_\ast$.
This method has been applied in \cite{khaneja2}-\cite{nimbalkar} assuming that one can apply a sequence of selective hard pulses on individual
qubits in the chain, and numerical solutions have been found for the 
time duration $\tau_\ast$ for an arbitrary couplings' ratio $K$. 

Here, however, we proceed in a different way.
In fact, from the variational principle and the action \Ref{eq-action}, we already have proof that the general time-optimal 
unitary evolution (up to an arbitrary target) must satisfy the QB equation \Ref{eq-fund}, and that such evolution is 
generated by the Hamiltonian \Ref{ising} with optimal magnetic field \Ref{bopt}. 
The global time-optimal solution depends upon the specific
boundary conditions and can be found by consequently determining the integration constants in \Ref{bopt}.
Therefore, we directly insert \Ref{ising} and \Ref{bopt} into 
$\langle O\rangle^\cdot =i\langle [H, O] \rangle$ and, defining the vector 
$\vec{x}:= (x_1, x_2, x_3, x_4, x_5, x_6, x_7, x_8)^t$, we obtain the system of equations:
\begin{align}
\frac{d\vec{x}(\tau)}{d\tau}=M(\tau)\vec{x}(\tau),
\label{system}
\end{align}
where the $8\times 8$ matrix: 
\begin{align}
 M(\tau):=2\left [ \begin{array}{cc}
 P(\tau)&Q\\
 Q&P(\tau)
  \end{array}\right ]
 \label{matrixM} 
\end{align}
depends on the skew symmetric, time-dependent $4\times 4$ matrix:
\begin{align}
 P(\tau):=\left [ \begin{array}{cccc}
 0&0&-1&0\\
  0&0&\hat B_0\sin\theta&-\hat B_z\\
 1&-\hat B_0\sin\theta&0&\hat B_0\cos\theta\\
 0&\hat B_z&-\hat B_0\cos\theta&0
  \end{array}\right ],
 \label{matrixP} 
\end{align}
(with $\theta(\tau)$ given below eq. \Ref{bopt}) and on the skew symmetric, constant $4\times 4$ matrix:
\begin{align}
 Q:=K\left [\begin{array}{cccc}
 0&0&0&0\\
  0&0&0&-1\\
 0&0&0&0\\
 0&1&0&0
  \end{array}\right ].
 \label{matrixQ} 
\end{align}
Our task is to solve the system \Ref{system} subject to the boundary conditions $x_1(0)=1, x_8(\tau_\ast)=1$, and to determine the values of the unknown integrals of the motion $\hat B_0, \hat B_z$ and $\hat \Omega$ in \Ref{bopt} such that the time duration of the transfer $\tau_\ast$ is minimal. We simplify the problem by introducing the 4-dimensional vectors 
$\vec{x}_+(\tau):=(x_1, x_2, x_3, x_4)^t$ and $\vec{x}_-(\tau):=(x_5, x_6, x_7, x_8)^t$, their combinations:
 $\vec{y}_\pm(\tau):=\vec{x}_+(\tau) \pm \vec{x}_-(\tau)$ and the $4\times 4$ matrices $M_\pm(\tau):=2[P(\tau)\pm Q]$.
 Then, eq. \Ref{system} decouples into the two sets of first order, linear differential equations:
\begin{align}
\frac{d\vec{y}_\pm(\tau)}{d\tau}=M_\pm(\tau)\vec{y}_\pm(\tau),
\label{system2}
\end{align}
whose solutions are:
 \begin{align}
\vec{y}_\pm(\tau)=\exp[A_\pm(\tau)]\vec{y}_\pm(0),
 \label{y}
 \end{align}
 where the $4\times 4$ matrix:
 \begin{align}
 A_\pm(\tau):=\int_0^\tau~M_\pm (s) ds = 2\left ( \begin{array}{cc}
 0&-R_\pm(\tau)\\
 \da R_\pm(\tau)&R_0(\tau)
  \end{array}\right ), 
 \label{matrixapm}
 \end{align}
and the $2\times 2$ matrices: 
 \begin{align}
 R_\pm(\tau)&:=\left ( \begin{array}{cc}
 \tau &0\\
 \frac{\hat B_0}{\hat \Omega}[\cos\theta(\tau)-\cos\theta_0]&(\hat B_z\pm K)\tau
  \end{array}\right ),
  \\
 R_0(\tau)&:=i\frac{\hat B_0}{\hat \Omega}[\sin\theta(\tau)-\sin\theta_0]\sigma_y.
 \label{matrixR} 
\end{align}

 For the transfer of coherence \Ref{transfer}, the initial and final boundary conditions can be rewritten 
 as $\vec{y}_+(0)=\vec{y}_-(0):=(1, 0, 0, 0)^t$ and $\vec{y}_+(\tau_\ast)=-\vec{y}_-(\tau_\ast):=(0, 0, 0, 1)^t$.
 After a long but straightforward analysis (for a sketch of the derivation see the Appendix), we finally find that the time-optimal transfer of coherence is characterized by the following integrals 
 of the motion.
The minimal time duration of the transfer is \footnote{Obtained for $m_0=n_0=0$, see the Appendix.}:
 \begin{align}
\tau_{\ast \mathrm{opt}}=\frac{\sqrt{3}}{4}\pi,
 \label{tauopt}
 \end{align}
while the optimal magnetic field (also using the energy constraint \Ref{eq-normH}) reads:
 \begin{align}
 \hat B_z&=0
 \\
 \hat B_0&=\pm K \sqrt{\hat \omega^2-2}
 \\
 \hat \Omega &=\pm \frac{4}{\pi}\sqrt{\hat \omega^2-2}
 \\
 \theta_0&=\frac{1}{2}[(2r+1)\pi \pm \sqrt{3}\sqrt{\hat \omega^2-2}]
 \label{b0bzomopt}
 \end{align}
where $r$ is an arbitrary integer and $\hat\omega  > 2$.
 It is worth noticing that the same time-optimal solution (with time duration and control as in \Ref{tauopt}-\Ref{b0bzomopt}) is also valid for for the transfer of coherence from the initial polarization
 $\sigma_x^1$ to the final operator $\sigma_y^1\sigma_x^2\sigma_z^3$ \footnote{Simply exchange the constants $b$ and $d$ everywhere in the Appendix.}.
 No transfer is possible, instead, from $\sigma_x^1$ to $\sigma_y^1\sigma_z^2\sigma_z^3$.
 
 \section{Discussion}

In this paper we described the most recent developments of the QB formalism.
We investigated the model of a trilinear Ising Hamiltonian with equal (or opposite) Ising couplings, subject to a local, time consuming control on the intermediate qubit and with a finite energy available.
The formalism of the QB \cite{pure} enables to find the exact, analytical form of the laws of the 
time-optimal control for the transfer from a single spin polarization
$\sigma_x^1$ to the final three-qubit coherence $\sigma_y^1\sigma_y^2\sigma_z^3$.
The analytical expression of the time duration of the transfer, equation \Ref{tauopt}, matches with a numerical result which
appeared in the previous literature \cite{khaneja2} for the same Ising interactions in the trilinear chain, but with only instantaneous local controls available \footnote{On the other hand, the QB imposes the finite energy condition eq. \Ref{eq-normH}, which is not assumed in \cite{khaneja2}.}.
It was proven here that this time of transfer is optimal.
An extension to the case of different couplings in the trilinear chain, to the situaiton where local controls are allowed on 
all the qubits in the chains, and to chains with more qubits and nonlinear topologies is in progress.
It is also important to consider the more general ansatz  where the total energy available is not a constant, but 
only bounded from above.

In the standard paradigm of time-optimal quantum computing (see, e.g., \cite{khaneja}) one-qubit unitary operations are assumed to have zero time cost. In more physical situations where this time cost cannot be neglected  
with respect to the other time-scales appearing in the quantum control problem, the neat geometrical methods of the standard paradigm cannot be immediately applied. The QB provides a natural formalism for this scenario.


\section*{ACKNOWLEDGEMENTS}
This research was partially supported by the MEXT of Japan, 
 under grant No. 09640341 (T.K.).
A.C. acknowledges the support from the MIUR of Italy under
the program "Rientro dei Cervelli".

\section*{APPENDIX}

The initial and final boundary conditions $\vec{y}_+(0)=\vec{y}_-(0):=(1, 0, 0, 0)^t$ and 
$\vec{y}_+(\tau_\ast)=-\vec{y}_-(\tau_\ast):=(0, 0, 0, 1)^t$ can be rewritten in terms of the matrix $A_\pm$ as: 
 \begin{align}
\left  [e^{A_\pm(\tau_\ast)}\right]_{11}&=\left  [e^{A_\pm(\tau_\ast)}\right]_{21}=\left  [e^{A_\pm(\tau_\ast)}\right]_{31}=0;
\nonumber \\
\left  [e^{A_\pm(\tau_\ast)}\right]_{41}&=\pm 1.
 \label{abound}
 \end{align}

Using the notation:
 \begin{align}
 a&:=2\tau_\ast,
\label{abcdin} \\
 c_\pm&:=(\hat B_z \pm K)a,
 \\
 b&:=2\frac{\hat B_0}{\hat\Omega}[\cos\theta(\tau_\ast)-\cos\theta_0] ,
\\
 d&:=2\frac{\hat B_0}{\hat\Omega}[\sin\theta(\tau_\ast)-\sin\theta_0], 
  \label{abcdout}
 \end{align}
 and introducing the quantities
 \begin{align}
X_\pm&:=a^2+b^2+c_\pm^2+d^2,
 \label{Xpm}
\\
\Delta_\pm&:= X_\pm^2-4a^2c_\pm^2,
 \label{deltapm}
\\
\sqrt{2} Z_\pm&:=\sqrt{X_+\pm \sqrt{\Delta_+}},
\label{Zpm}
\\
\sqrt{2} W_\pm&:=\sqrt{X_-\pm \sqrt{\Delta_-}},
 \label{Wpm}
 \\
 C^Z_\pm&:=\cos Z_\pm;~~~ C^W_\pm :=\cos W_\pm,
 \label{coszw}
 \\
 S^Z_\pm&:=\frac{\sin Z_\pm}{Z_\pm}; ~~~ S^W_\pm :=\frac{\sin W_\pm}{W_\pm}, 
 \label{sinzw}
  \end{align}
we can explicitly write the boundary conditions \Ref{abound} (in the 
order of components $(11)_\pm ; (31)_\pm ; (21)_\pm ; (41)_\pm$) as: 
 \begin{align}
 [X_+-2a^2-\sqrt{\Delta_+}&]C^Z_+=[X_+-2a^2+\sqrt{\Delta_+}]C^Z_-
\label{11p}
\\
 [X_--2a^2-\sqrt{\Delta_-}]&C^W_+=[X_--2a^2+\sqrt{\Delta_-}]C^W_-
\label{11m}
\\
[X_+-2c_+^2+\sqrt{\Delta_+}]&S^Z_+=[X_+-2c_+^2-\sqrt{\Delta_+}]S^Z_-
\label {31p}
\\
[X_--2c_-^2+\sqrt{\Delta_-}]&S^W_+=[X_--2c_-^2-\sqrt{\Delta_-}]S^W_-
\label{31m}
\\
b(C^Z_+-C^Z_-)&=c_+ d (S^Z_+ -S^Z_-),
\label{21p}
\\
b(C^W_+-C^W_-)&=c_- d (S^W_+ -S^W_-),
\label{21m}
\\
d(C^Z_+-C^Z_-)&=-c_+ b (S^Z_+ -S^Z_-)+\frac{\sqrt{\Delta_+}}{a},
\label{41p}
\\
d(C^W_+-C^W_-)&=-c_- b (S^W_+ -S^W_-)-\frac{\sqrt{\Delta_-}}{a}.
\label{41m}
 \end{align}
 An elementary  analysis of the system of equations \Ref{11p}-\Ref{41m} leads to find out that its only
 solution is possible for the values of the parameters $d=0$ and $b\not =0$.
 We now sketch how we get to the complete, analytical form of the solutions for the time-optimal transfer of coherence.
 First, for $d=0$ but $b\not =0$, and from eqs. \Ref{21p}-\Ref{21m}, one finds that
$Z_-=-Z_+ +2\pi p$ and $W_+=-W_- +2\pi q$, 
where $p$ and $q$ are arbitrary and (due to the definitions \Ref{Zpm}-\Ref{Wpm}) nonnegative integers \footnote{Of course, also $Z_-=Z_++2\pi p$ and $W_+=W_-+2\pi q$ in principle are also solutions for \Ref{21p}-\Ref{21m}. However, it can be easily shown that for these values the rest of equations have no solution.}. 
 Then, substituting the latter relations into \Ref{11p}-\Ref{11m}, we find that the only possible choice \footnote{We exclude the trivial case $a=b=c_\pm=0$, which would give a zero time duration of the transfer.} is 
$Z_+=\frac{\pi}{2}(2n+1)$ and $W_-=\frac{\pi}{2}(2m+1)$,
 where again $m$ ad $n$ are arbitrary, nonnegative integers. 
 
 We can now  eliminate $Z_\pm$ and $W_\pm$ from the square of eqs. \Ref{Zpm}-\Ref{Wpm} to get:
\begin{align}
\sqrt{\Delta_+}&=2\pi^2p[(2n+1)-2p],
\label{deltap}
\\
\sqrt{\Delta_-}&=-2\pi^2q[(2m+1)-2q],
 \label{deltam}
\\
 a^2+b^2+c_+^2&=\frac{\pi^2}{2}[(2n+1)^2-4p(2n+1)+8p^2],
 \label{xp}
\\
 a^2+b^2+c_-^2&=\frac{\pi^2}{2}[(2m+1)^2-4q(2m+1)+8q^2].
 \label{xm}
 \end{align}
In other words, we have replaced the original system of eqs. \Ref{11p}-\Ref{41m} with eqs. \Ref{deltap}-\Ref{xm} (instead of eqs. \Ref{11p}-\Ref{11m} and eqs. \Ref{21p}-\Ref{21m}) and (after substitution of $Z_-, W_-$ into eqs. \Ref{31p}-\Ref{31m} and \Ref{41p}-\Ref{41m}) with equations:
  \begin{align}
 \sqrt{\Delta_+}&=(a^2+b^2-c_+^2)\frac{2p}{[(2n+1)-2p]},
\label{deltap1}
\\
 \sqrt{\Delta_-}&=-(a^2+b^2-c_-^2)\frac{2q}{[(2m+1)-2q]}, 
 \label{deltam1}
\\
 abc_+&=(-1)^{n+1}(a^2+b^2-c_+^2)\frac{\pi (2n+1)[(2n+1)-4p]}{4[(2n+1)-2p]}, 
 \label{abcp}
\\
 abc_-&=(-1)^{m}(a^2+b^2-c_-^2)\frac{\pi (2m+1)[(2m+1)-4q]}{4[(2m+1)-2q]}. 
 \label{abcm}
 \end{align}
After a short algebra, the solution to eqs. \Ref{deltap}-\Ref{abcm} is given by 
 $2p=2q=m+n+1$ and the following constants:
 \begin{align}
a&=\pm c_\pm=\frac{\pi}{2}\sqrt{(2m+1)(2n+1)},
\label{acopt}
\\
b&=-\pi K(n-m),
 \label{abopt}
 \end{align}
 with $|K|=1$ and where the integers $m$ and $n$ have opposite parity ($m=2m_0$ and $n=2n_0+1$) and $n>n$
 (i.e. $n_0\geq m_0\geq 0$).
Finally, we can use the results \Ref{acopt}-\Ref{abopt} 
to invert \Ref{abcdin}-\Ref{abcdout} in favor of the physical quantities $\tau_\ast, \hat B_0,
\hat B_z, \hat \Omega$ and $\theta_0$. Minimization of $\tau_\ast$ with respect to $m$ and $n$  gives 
eqs. \Ref{tauopt}-\Ref{b0bzomopt} of the main text.

\bibliographystyle{alpha}

\end{document}